\documentclass[fleqn,1p]{elsarticle}
\usepackage{graphicx}
\usepackage{natbib}
\usepackage{amsmath,url}
\usepackage{multirow}
\usepackage{tabularx}
\usepackage{hhline}
\usepackage{rotating}
\usepackage{algorithm}
\usepackage{algorithmicx,algpseudocode}
\usepackage{color, colortbl}
\usepackage{tabu}
\usepackage{xcolor}
\definecolor{Gray}{gray}{0.9}
\usepackage{amssymb}
\usepackage{array}
\newcolumntype{L}[1]{>{\raggedright\let\newline\\\arraybackslash\hspace{0pt}}m{#1}}
\newcolumntype{C}[1]{>{\centering\let\newline\\\arraybackslash\hspace{0pt}}m{#1}}
\newcolumntype{R}[1]{>{\raggedleft\let\newline\\\arraybackslash\hspace{0pt}}m{#1}}
\usepackage{color,soul}
\journal{Digital Signal Processing}
\begin{document}
\begin{frontmatter}
\title{Optimization of data-driven filterbank for automatic speaker verification}
\author[mysecondaryaddress]{Susanta Sarangi\corref{corr1}}
\ead{sksarangi@ece.iitkgp.ac.in}
\address[mysecondaryaddress]{Department of Electronics \& Electrical Communication Engineering, \\ Indian Institute of Technology, India-721302, Kharagpur, India }
\author[mymainaddress]{Md Sahidullah}
\ead{md.sahidullah@inria.fr}
\author[mysecondaryaddress]{Goutam Saha}
\ead{gsaha@ece.iitkgp.ernet.in}
\address[mymainaddress]{Universit\'{e} de Lorraine, CNRS, Inria, LORIA, F-54000, Nancy, France}
\cortext[corr1]{Corresponding author}
\begin{abstract}
\par
Most of the speech processing applications use triangular filters spaced in mel-scale for feature extraction. In this paper, we propose a new data-driven filter design method which optimizes filter parameters from a given speech data. First, we introduce a frame-selection based approach for developing speech-signal-based frequency warping scale. Then, we propose a new method for computing the filter frequency responses by using \emph{principal component analysis} (PCA). The main advantage of the proposed method over the recently introduced deep learning based methods is that it requires very limited amount of unlabeled speech-data. We demonstrate that the proposed filterbank has more speaker discriminative power than commonly used mel filterbank as well as existing data-driven filterbank. We conduct \emph{automatic speaker verification} (ASV) experiments with different corpora using various classifier back-ends. We show that the acoustic features created with proposed filterbank are better than existing \emph{mel-frequency cepstral coefficients} (MFCCs) and \emph{speech-signal-based frequency cepstral coefficients} (SFCCs) in most cases. In the experiments with VoxCeleb1 and popular i-vector back-end, we observe 9.75\% relative improvement in equal error rate (EER) over MFCCs. Similarly, the relative improvement is 4.43\% with recently introduced x-vector system. We obtain further improvement using fusion of the proposed method with standard MFCC-based approach.
\end{abstract}
\end{frontmatter}
\section{Introduction}
Speech is a short-term stationary signal~\citep{quatieri2006discrete} which contains information related to the spoken content, speaker's identity, speaker's emotion, spoken language, etc. The speaker recognition technology recognizes persons from their speech~\citep{kinnunen2010overview1}. The \emph{automatic speaker verification} (ASV) is one of the important tasks in speaker recognition where two voice signals are compared by machine for deciding whether they are produced by the same speaker or not. The ASV technology finds its application in voice biometrics for authentication tasks in both logical and physical access scenarios~\citep{sahidullah2019introduction, markowitz2000voice} and also help in the judicial system to compare an unknown speaker's voice with a known suspect's voice~\citep{dumas1990voice, campbell2009forensic}. The performance of the ASV system is reliable in \emph{controlled conditions}; however, in the real-world situations, the performance is considerably degraded due to the variations in \emph{intrinsic factors} (speaker's emotion, health, age, etc.) and \emph{extrinsic factors} (background noise, channel, room impulse response, etc.)~\citep{nagrani2017voxceleb1}. To achieve a good performance in practical applications, the ASV system should be robust against these unwanted variations.
\par
A typical ASV system consists of three main modules: \emph{frame-level feature extractor}, \emph{segment-level feature (embedding) extractor}, and \emph{classifier}. The frame-level feature extraction unit converts  raw speech waveform into a sequence of acoustic feature vectors~\citep{kinnunen2010overview1, campbell1997speaker}. Most of the ASV studies use short-term spectral features which are based on the knowledge of speech production and perception model. Some studies use high-level features as complementary information which represent other speaking characteristics such as, speaking rate and pronunciation style~\citep{ayoub2016gammatone}. The classifier module further parameterizes the features into statistical models~\citep{kinnunen2010overview1}. For efficient use of the ASV systems in different real-world applications, we need a feature extraction method which should be robust to unwanted variations in the speech signal and computationally inexpensive~\citep{kinnunen2010overview1}. Improving the robustness of acoustic feature usually reduces the effort from classifier for improving the ASV system performance. The scope of this work is limited to the development of a new robust feature extraction algorithm for real-world applications.

Among all the existing cepstral features, \emph{mel-frequency cepstral coefficients} (MFCCs) are the most popular and widely used for the ASV as well as other speech processing tasks such as automatic speech recognition~\citep{benzeghiba2007automatic}, speaker diarization~\citep{anguera2012speaker}, spoofing countermeasures~\citep{wu2015spoofing}, etc. The recently introduced \emph{x-vector} based ASV system, which drew attention in previous NIST speaker recognition evaluations~\citep{NIST,lee2019i4u,NISTSRE2019}, also uses MFCCs as acoustic features. The MFCC computation process involves mel scale integration followed by logarithmic compression and \emph{discrete cosine transform} (DCT). The MFCCs are very popular for the following reasons. First, the computation process~\textcolor{black}{utilizes} mel filterbank analysis, which is partially inspired by the processing of the audio signal by the human auditory system. Second, the computation process involves \emph{fast Fourier transform} (FFT) and matrix multiplication which makes it more computationally efficient compared to other methods such as \emph{linear prediction cepstral coefficients} (LPCCs) or \emph{line spectral frequencies} (LSFs)~\cite{RabinerFundamentals1993}. Third, MFCCs are also suitable for different feature-level compensation methods such as \emph{relative spectral} (RASTA) processing~\cite{hermansky1994rasta}, \emph{cepstral mean and variance normalization} (CMVN), and \emph{feature warping}~\cite{Pelecanos2001}. Though the MFCCs are relatively more robust compared to other cepstral features such as \emph{linear frequency cepstral coefficients} (LFCCs) or LPCCs, the ASV performance with MFCCs are severely degraded in real-world conditions due to the mismatch of acoustic conditions in enrollment (or speaker registration) and verification (or speaker authentication) phase~\citep{sahidullah2012design, paliwal2009speech}. To overcome some of the shortcomings of MFCCs, various acoustic features like \emph{frequency domain linear prediction} (FDLP)~\citep{ganapathy2012signal}, \emph{cochlear frequency cepstral coefficients} (CFCCs)~\citep{li2011auditory}, \emph{power-\textcolor{black}{normalized} cepstral coefficients} (PNCCs)~\citep{kim2016power}, \emph{mean Hilbert envelope coefficients} (MHECs)~\citep{SADJADI2015138}, \emph{Gammatone frequency cepstral coefficients} (GFCCs)~\citep{zhao2012casa}, \emph{constant-Q cepstral coefficients} (CQCCs)~\citep{todisco2016articulation}, \emph{time-varying linear prediction} (TVLP)~\citep{vestman2018speaker}, and \emph{locally-\textcolor{black}{normalized} cepstral coefficients} (LNCCs)~\citep{poblete2015perceptually} were proposed. 
All these features even though achieve better performance in noisy condition, they require a large number of user-defined parameters. These parameters further need to be manually tuned for different environmental conditions. The overall process seems to be difficult for a system-developer. Also, improving feature robustness beyond a certain level is extremely difficult, especially for a wide range of degradation~\citep{ganapathy2012signal,SADJADI2015138}. Besides, most of those features are also computationally more expensive than MFCCs. The MFCCs, on the other hand, have lesser number of free parameters. This study develops a data-driven feature extraction method which follows the same principle as MFCC but derives the parameters from the speech data itself. Unlike the feature extraction methods discussed before, which require ``hand-crafted" parameters, the feature extraction method with parameters computed in a \emph{data-driven} procedure reduces the effort needed for manual fine-tuning. The data-driven methods also show the improvement in robustness when large corpora are used in training strong discriminative models~\citep{hansen2015speaker}.

\textcolor{black}{The data-driven acoustic feature extraction methods use speech data to compute the parameters of the feature extraction algorithm. We classify those methods into two broad categories. One of them uses discriminative approaches such as the \emph{artificial neural network} (ANN) or \emph{linear discriminant analysis} (LDA). These methods require labeled speech data. The other type does not apply the discriminative approach but utilizes some speech science knowledge during parameter estimation. In other words, they learn the feature extraction parameters directly from the speech data without using any class label information. Some of the popular data-driven speech feature extraction methods are discussed in Table~\ref{Literature}. Most of the methods are discriminative in nature, and they are generally investigated for automatic speech recognition (ASR) tasks. In ASV research, data-driven feature extraction methods have drawn relatively less attention~\citep{ravanelli2018speaker}.}

\begin{table}[t!]
\renewcommand{\arraystretch}{1.2}
\caption{Selected works on data-driven feature extraction methods for various speech applications (ASR: Automatic speech recognition, ASV: Automatic speaker verification, SAD: Speech activity detection).}
 \centering
 \begin{footnotesize}
\begin{tabular}{|c|>{\arraybackslash}p{10cm}|c|}
\hline
\textbf{Work} & \textbf{Methodology} & \textbf{Task}\\
\hline
 \multirow{2}{*}{\citep{hermansky1999temporal}}& Neural network is trained with speech features of larger temporal context and used to create data-driven features called TempoRAl Patterns (TRAPs).&\multirow{2}{*}{ASR}\\
 \hline
 \multirow{2}{*}{\citep{malayath2000data}}& This work investigates data-driven temporal filter with oriented principal component analysis (OPCA) that reduces channel variability.&\multirow{2}{*}{ASV}\\
\hline
 \multirow{2}{*}{\citep{burget2001data}}& The filterbank is derived from phonetically labeled speech data using LDA.&\multirow{2}{*}{ASR}\\
\hline
 \multirow{2}{*}{\citep{malayath2003data}}& Data-driven LDA is applied on the logarithmic critical-band power spectrum of speech.&\multirow{2}{*}{ASR}\\
\hline
 \multirow{5}{*}{\citep{hermansky2003trap}}& This method uses TRAP followed by TANDEM. The TRAP estimator provides multiple evidences in terms of posterior probabilities from frequency-localized overlapping time-frequency regions of speech signal computed with the help of data-driven transformation of contextual information. Next, TANDEM converts the frequency-\textcolor{black}{localized} evidences to features.&\multirow{5}{*}{ASR}\\
 \hline
 \multirow{2}{*}{\citep{hung2006optimization}}& Data-driven temporal filters are designed using PCA, LDA and minimum classification error (MCE) framework.&\multirow{2}{*}{ASR}\\
\hline
 \multirow{2}{*}{\citep{el2006using}}& Speech segments are created using a data-driven and automatic language independent speech processing (ALISP).&\multirow{2}{*}{ASV}\\
\hline
\multirow{2}{*}{\citep{lu2008investigation}}& This work uses F-ratio to adjust the center and edge frequencies of the filterbank and the F-ratio is computed for speaker separability.&\multirow{2}{*}{ASV}\\
\hline
 \multirow{2}{*}{\citep{paliwal2009speech}}& Data-driven frequency warping is obtained by dividing the long-term average spectrum (LTAS) into subbands of equal energies.&\multirow{2}{*}{ASR}\\
\hline
 \multirow{2}{*}{\citep{thomas2012acoustic}}& A multi-layer perceptron (MLP) is trained to classify speech and non-speech frames. The outputs of the MLP are used as posterior features.&\multirow{2}{*}{SAD}\\
\hline
 \multirow{3}{*}{\citep{sainath2015learning}}& Combination of convolutional neural network (CNN) and long short-term memory (LSTM) is used to learn neural network parameters to classify the context-dependent state labels.&\multirow{3}{*}{ASR}\\
\hline
 \multirow{2}{*}{\citep{hoshen2015speech}}& CNN is used to learn time-domain filter parameters and the network is trained to classify the context-dependent state labels.&\multirow{2}{*}{ASR}\\
\hline
 \multirow{2}{*}{\citep{7563327}}& The filterbank is learned in an unsupervised manner using convolutional restricted Boltzmann machine (ConvRBM) with clean and noisy audio data.&\multirow{2}{*}{ASR}\\
\hline
 \multirow{2}{*}{\citep{seki2017deep}}& The triangular mel filter is approximated using Gaussian function and the parameters of this pseudo filter are learned using DNN. &\multirow{2}{*}{ASR}\\
\hline
 \multirow{3}{*}{\citep{zeghidour2018learning}}& Computational steps of mel-frequency spectral coefficients (MFSCs) are implemented with neural network where the parameters are learned using convolution layers with a goal of maximizing phone recognition accuracy.&\multirow{3}{*}{ASR}\\
\hline
 \multirow{4}{*}{\citep{ravanelli2018speaker}}& A CNN-based architecture SincNet is introduced which learns the lower and upper cut-off frequencies of the subband filters. Each filter is approximated with the help of a pair of Sinc functions in time-domain and its parameters are tuned  by maximizing speaker classification accuracy.&\multirow{4}{*}{ASV}\\
\hline
\end{tabular}
\end{footnotesize}
\label{Literature}
\end{table}

\textcolor{black}{In this work, we perform detailed analysis of a data-driven feature extraction method for ASV which utilizes only audio-data for computing the desired parameters, in contrast to most of the data-driven techniques that require additional metadata such as speech (\emph{e.g.}, phoneme) or speaker information. We select \emph{speech-signal-based frequency cepstral coefficient} (SFCC), and this feature has demonstrated promising performance in speech and speaker recognition applications~\citep{paliwal2009speech, sarangi2012novel}. The method is also very similar to MFCC; however, in contrast to MFCC which applies handcrafted mel scale, SFCC utilizes a \emph{frequency warping} scale that is computed by a data-driven approach. Since the filterbank parameters are computed prior to the feature extraction step, its effective computational time is same as that of MFCCs, and thus considerably lower than other recently proposed features such as FDLP, MHEC or CQCC. The current study extends our preliminary study~\citep{sarangi2012novel} which introduced the basic data-driven frequency warping~\citep{paliwal2009speech} in speaker recognition. In this work, we further improve this method by optimizing the scale and by computing the other parameters in a data-driven manner. By performing separability analysis with F-ratio, we have demonstrated that the proposed features are more speaker discriminative than standard MFCCs. Our ASV experiments conducted with different ASV systems agree with this analysis. The major contributions of this work are summarized below.}

\begin{itemize}
    \item \textcolor{black}{We improve the basic data-driven scale with frame selection. With comprehensive analysis and experimental results, we demonstrate that selective use of speech-frames helps to create more reliable frequency warping scale.}
 
    \item \textcolor{black}{We introduce a data-driven way for computing filter responses as an alternative to the auditory motivated triangular filters. Our proposed method computes the filterbank response in an unsupervised way with a smaller amount of speech data in contrast to the discriminative approaches that require class labels and a larger amount of speech data.}
    
    \item \textcolor{black}{We evaluate the proposed features with a state-of-the-art x-vector based ASV system which currently utilizes either MFCCs or log-mel energy features.} 
\end{itemize}


The rest of the paper is organized as follows. Section~\ref{Section:Cepstral} explains the baseline cepstral feature extraction methods for both mel and data-driven scale. The next section presents the proposed method for improving the data-driven scale. We propose the data-driven approach of computing filter responses in Section~\ref{Section:Optimization of filter shape}. We discuss the experimental setup in Section~\ref{Section:Experimental setup}, and we show the results in Section~\ref{Section:Results and discussion}. Finally, we conclude in Section~\ref{Section:Conclusion} with a discussion on limitations of this study and possible future directions.

\section{Cepstral features based on filterbank}\label{Section:Cepstral}

\begin{figure}[h]
\centering
\includegraphics[width=1\textwidth]{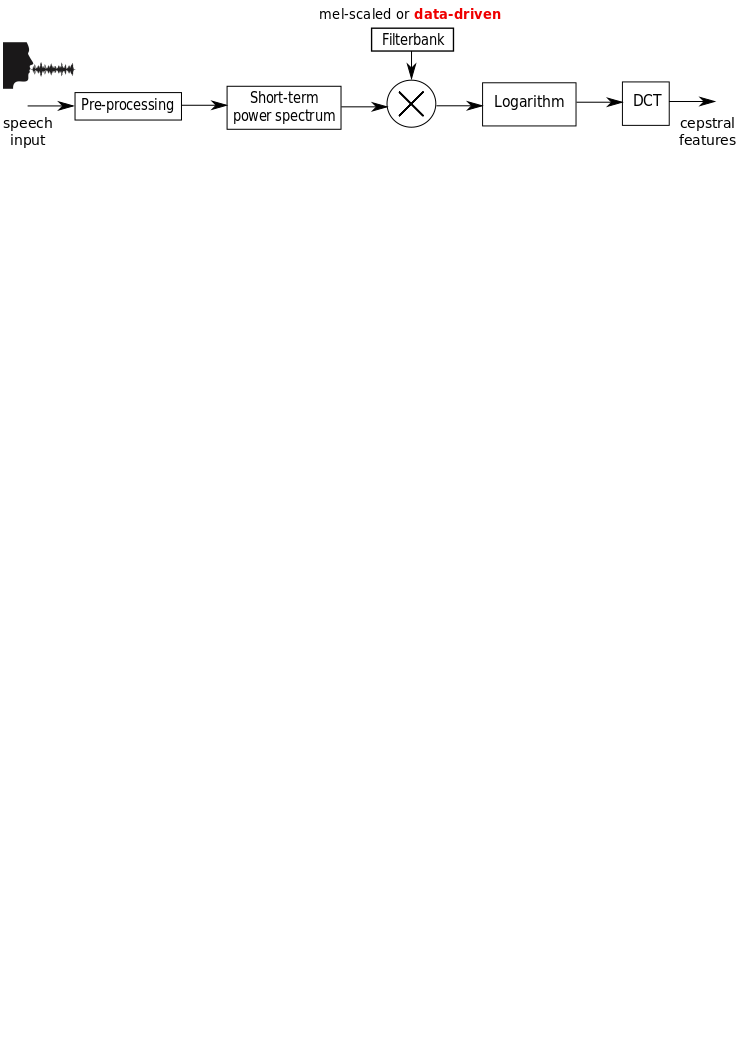}\\
\vspace*{-16.5cm}
\caption{Block diagram of a typical cepstral feature extraction method.}\label{fig:1}
\centering
\end{figure}

A general block diagram of cepstral feature extraction methods using a filterbank is shown in Fig.~\ref{fig:1}. After pre-processing steps such as framing and windowing, the short-term power spectrum of speech frames is multiplied with a filterbank frequency response. Then, cepstral features are created by performing DCT on log-energies of filterbank output. In MFCC computation, we place the triangular-shaped filters in the mel scale. However, for SFCCs, triangular filters are placed in data-driven speech-signal-based scale. In the following sub-sections, we briefly describe these two feature extraction methods.

\subsection{MFCCs: fixed mel scale}
The MFCC feature extraction scheme introduced in~\citep{davis1980comparison} provides a straightforward way to compute cepstral features. Since then, it had been the state-of-the-art in different speech-based applications including speaker recognition~\citep{8461375}. It uses the mel scale~\citep{Stevens} based triangular filterbank for the creation of cepstral features. There are several alternatives to mel scale representations~\citep{ganchev2005comparative}. The most commonly used equation to convert linear frequency $\mathit{f}$ to mel frequency $\mathit{f}_{mel}$ is

\begin{equation}\label{eqpia1}
f_{mel}=2595\log_{10}\left(1+\frac{f}{700}\right).  
\end{equation}

In the mel scale domain, the frequency axis is divided into equidistant points. By considering those points as filter edge-frequencies, the filters are placed by keeping $50\%$ overlap with the adjacent one~\citep{Sandipan1}. In the MFCC computation step, the pre-emphasized speech signal is first segmented into frames of $20$-$30$~ms typically with an overlap of $50\%$. After that, we perform the short-time Fourier transform (STFT) of the speech frames. Then, we compute filterbank energies by using mel filterbank. Finally, DCT is performed over logarithm of filterbank energies to get cepstral features. The detailed procedure to compute MFCCs can be found in~\citep{sahidullah2012design, Ganchev}.

\subsection{SFCCs: data-driven scale}
The SFCC is a data-driven cepstral feature extraction method which computes the non-linear scale from the training speech. This scale was initially proposed for speech recognition~\citep{paliwal2009speech} and later has been successfully applied in speaker recognition~\citep{sarangi2012novel}. The SFCC extraction method replaces the mel scale with a data-driven scale, and the rest of the process is the same as the MFCC computation process. The following paragraph describes the steps required to get data-driven scale.

The scale computation involves the computation of long-term average power spectrum (LTAS) of speech data. The LTAS per speech utterance is computed first by averaging the short-term power spectrum over all the frames in the utterance. \textcolor{black}{Then, average LTAS is computed over all the speech utterances present in a corpus for computating the scale}. In the next step, the logarithm of LTAS is divided into equal area intervals to compute the filter edge frequencies.

The derivation of the speech-based data-driven scale is described in the following steps.

\textbf{1. Computation of LTAS}: Let $\mathit{u}$ be a speech utterance of $\mathit{N}_l$ frames. Its LTAS is expressed as,

\begin{equation}\label{eqpia2}
P[k]=\frac{1}{N_l}\sum_{i=1}^{N_l}X_i[k],
\end{equation}

where $X_i[k]$ is the energy spectrum and $k$ is the index of frequency bin.

\textbf{2. Computation of average LTAS}: The average LTAS is computed as the ensemble average of LTAS of all speech utterances in a corpus, and it is defined as,

\begin{equation}\label{eqpia3}
\bar{P}[k]=\frac{1}{N_s}\sum_{i=1}^{N_s}P[k],
\end{equation}

where $N_s$ is the total number of speech utterances in a corpus.

\textbf{3. Computation of cumulative log power spectrum}: Now, if we want to place $\mathit{Q}$ filters, we divide the $\log \bar{P}[k]$ into frequency subbands of $\mathit{Q}$ equal areas. We compute the area of the $j$-th band as,

\begin{equation}\label{eqpia4}
A_j=\sum_{k=k_l^j}^{k_h^j}\log\bar{P}[k],
\end{equation}

where $j=1,2,3,...,\mathit{Q}$. Here $k_l^j$ and $k_h^j$ are the lower and upper band for $j$-th filter and they are selected in such a manner that $A_1=A_2=A_3=\ldots=A_Q$. We also consider the lower edge frequency of the first filter as $0$~Hz and the higher edge frequency of the last filter as the Nyquist frequency. In practice, it is not possible to get $A_i$s exactly equal in numerical values, and they are made approximately equal.

\textbf{4. Computation of warping scale}: Finally, the scale is computed by interpolating the filterbank center frequencies to their mapped values which are obtained with the help of the following equation~\citep{paliwal2009speech},

\begin{equation}\label{eqpia5}
W\left[\frac{k_l^j+k_h^{j}}{2}\right]=\frac{j}{Q},
\end{equation}
where $j=1,2,3,..., \mathit{Q}$.

Eq.~(\ref{eqpia5}) gives the required frequency points to design filters in the filterbank structure. The cepstral features computed with this scale is referred to as SFCCs. This scale used in SFCC computation is shown in Fig.~\ref{fig:3} along with standard mel, ERB, and Bark scale as well as the scale proposed in Section~\ref{Section3}.

\textcolor{black}{To compute this scale, we do not require speaker labels for the corpus, unlike most of the methods listed in Table~\ref{Literature}}. During this scale computation, all the speech frames are used which are selected by a speech activity detection (SAD) method. This includes all types of speech frames showing different spectral characteristics; however, we do not necessarily need the entire speech corpus as LTAS can be obtained with a small subset of available data. In the next section, we consider a frame selection technique to select useful frames for better ASV performance.

\section{Data-driven frequency warping using selected frames}\label{Section3}

The frame selection strategy is used in speaker recognition task for fast implementation in real-time application~\citep{kinnunen2006real, sarkar2010real}. In this work, we select a subset of speech frames for developing warping scale. 

The conventional mel scale is a psychophysical scale for pitch perception. This experimental scale was first formulated with the help of a perceptual test by playing tones of fixed frequency to the listeners~\citep{Stevens}. The participants were asked to tune the frequency of another adjustable tone according to the perceived half-value of the fixed tone. All the tones were played at a constant loudness of $60$~dB. The scale was formulated by fitting a curve that maps the numerical value of linear frequency to the perceived value.

We note that the mel scale development is originally a subjective method which might be biased to the selected listeners~\citep{Greenwood}. Therefore, instead of subjective criterion, in data-driven method, we replace human being with the objective criterion of equal energy of voice signal. During this process, we consider all types of speech data irrespective of the voice production mechanism. This crude selection of speech frames include unvoiced speech frames created with random noise passing through a narrow constriction of the vocal tract. This unvoiced frames have no harmonic structure and closely resemble the uniform distribution of noise spectra~\citep{quatieri2006discrete}. Therefore, we propose to select only the voiced frames having pitch information for our data-driven scale formulation process.

\begin{figure}[t!]
\centering
\includegraphics[width=1\textwidth]{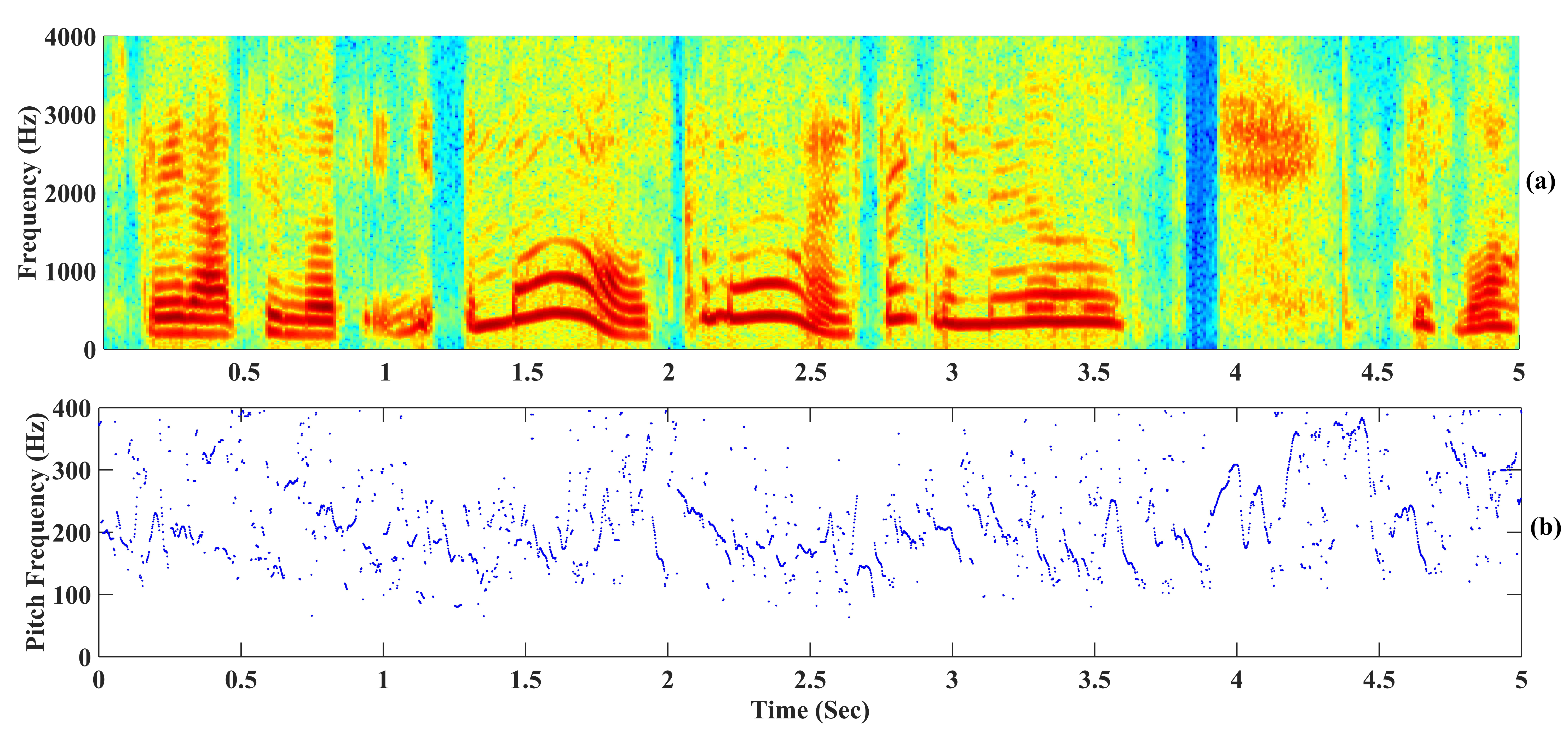}\\
\caption{\textcolor{black}{Illustration of (a)~spectrogram and (b)~pitch contour of a speech signal taken from NIST SRE $2001$ speech corpus. We compute the pitch values using \emph{the pitch estimation filter robust to high levels of noise} (PEFAC) method as studied in~\citep{gonzalez2014pefac}.}}\label{fig:2}
\centering
\end{figure}

\begin{figure}[t!]
\centering
\includegraphics[width=.98\textwidth]{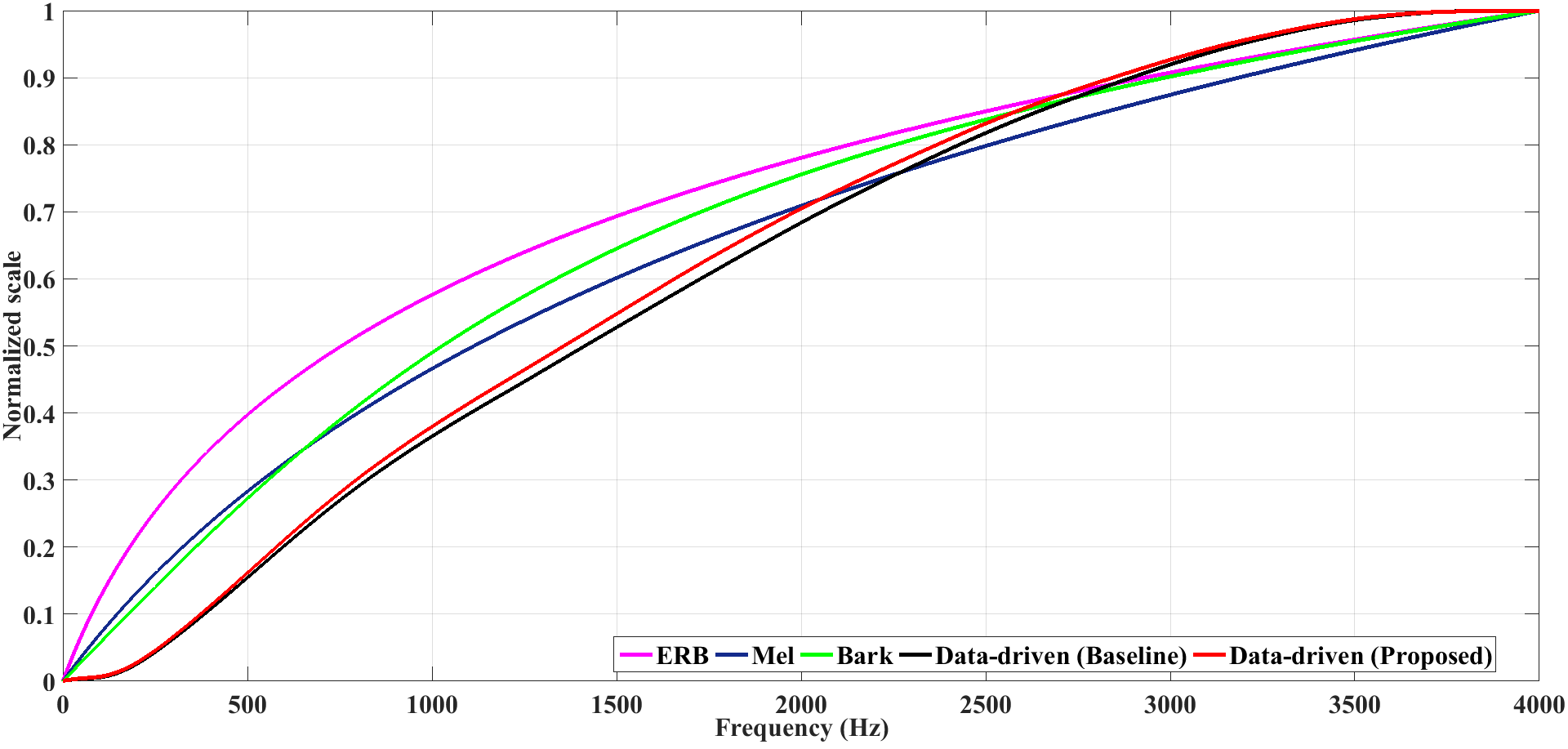}
\caption{The frequency warping function for Mel, ERB, Bark, SFCC, and proposed scale.}\label{fig:3}
\centering
\end{figure}

Fig.~\ref{fig:2} shows the spectrogram and pitch contour of the speech signal, which is taken from the NIST SRE $2001$ corpus.  

Fig.~\ref{fig:3} shows the~\textcolor{black}{normalized} plot of both auditory and data-driven scales. \textcolor{black}{We observe that the data-driven scales have lower resolution at the both ends of the frequency band, and higher resolution everywhere else. This is expected as the speech files for NIST SRE $2001$ are collected over telephone channel with a frequency range of $300-3700$~Hz. Therefore, we hypothesize that the filterbank placed according to the newly derived scale will help to capture more relevant information than mel filterbank or standard data-driven filterbank.}

\section{Computation of data-driven filter shape using PCA}\label{Section:Optimization of filter shape}
\par
In traditional MFCCs and SFCCs, we use filters with triangular-shaped frequency responses which closely approximate the auditory filters in cochlea. Other shapes like Gaussian~\citep{chakroborty2010feature} and trapezoidal~\citep{hermansky1990perceptual} are also used in speech feature extraction process. The shape of the filters in filterbank assigns non-uniform weights to the subband frequencies.

In this work, the idea is to design the subband filter response so that the output of this filter, computed as the energy, will represent the subband frequency components in a most effective way. In other words, we need to reduce the dimension of the subband frequency components to a single data point. We employ \emph{principal component analysis} (PCA) which is appropriate for finding the most ``expressive" representation of the data~\citep{bishop2006pattern}. Previously, PCA is applied to design the data-driven filterbank with mel scale for robust speech recognition~\cite{hung2006optimization, lee2001improved}. We propose to apply PCA on the log-power spectrum of each frequency band separately for constructing the filters. The PCA basis with highest eigenvalue, known as ``first basis", is used to create the filter frequency response. Since speech signal is a highly correlated process~\cite{0888Jayant1984}, the subband covariance matrix will be positive. Hence all the elements of its eigenvector with highest eigenvalue, i.e., the first basis of PCA, will be non-negative according to the \emph{Perron-Frobenius} theorem~\citep{johnson1985matrix}. The steps to find the filter shape are~\textcolor{black}{summarized} below:

\begin{figure}[t!]
\centering
\includegraphics[width=1\textwidth]{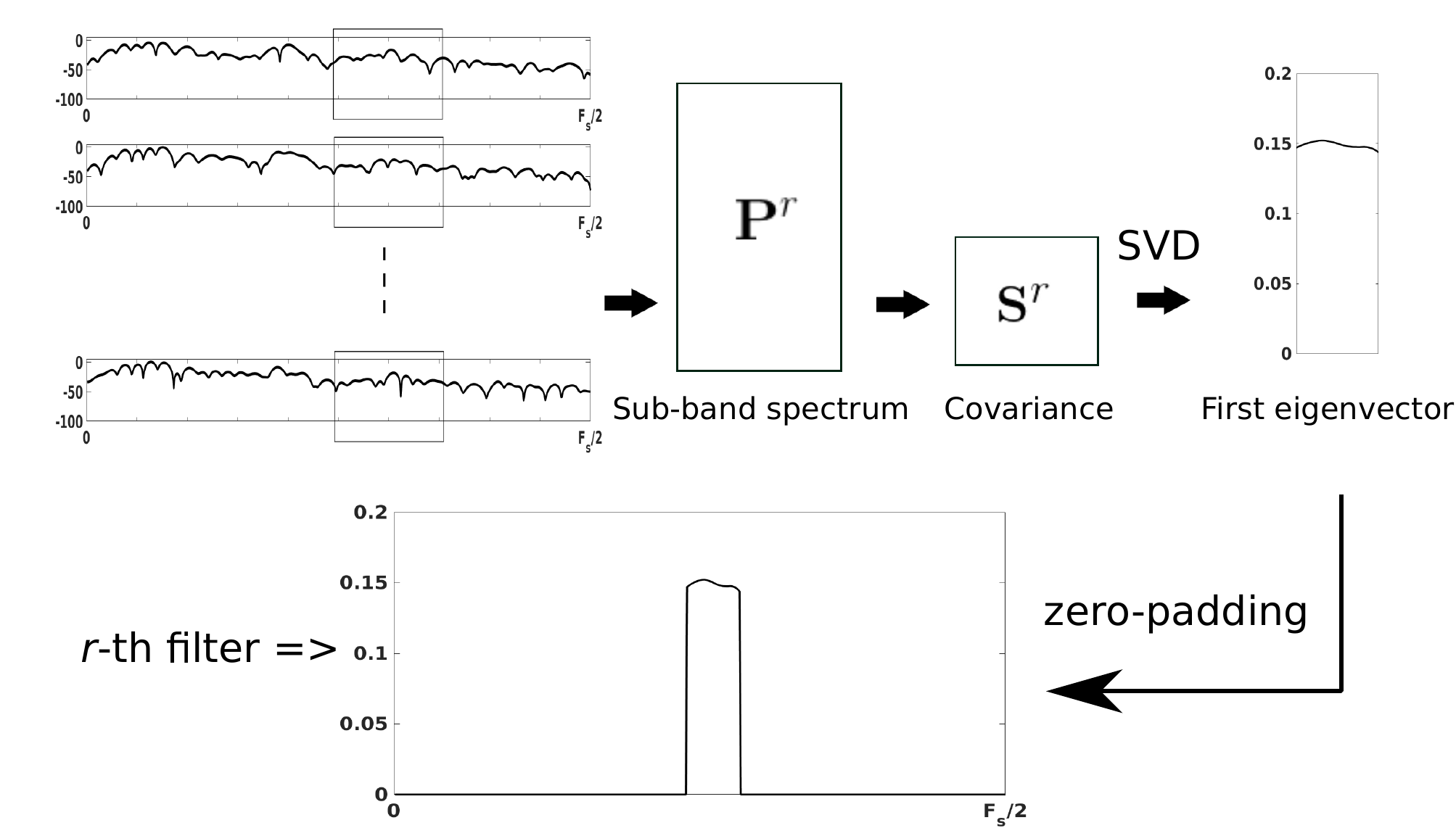}
\caption{Figure showing proposed data-driven method for computing frequency response of a filter. Here the matrix $\mathbf{P}^{r}$ contains log power spectrum of all the frames corresponding to the $r$-th subband.}\label{fig:4} 
\centering
\end{figure}

\textbf{1. Computation of subband covariance matrix}: Let $P_{i}^{r}[k]$ be the log power spectrum of $k$-th frequency component for $r$-th subband and $i$-th speech frame. Then the subband covariance matrix corresponding to the $r$-th subband is given as:

\begin{equation}\label{eqpia6}
\mathbf{S}^{r}=\frac{1}{N_{f}-1}\sum_{i=1}^{k_{r}}(P_{i}^{r}[k]-\bar{m}[k])(P_{i}^{r}[k]-\bar{m}[k])^{\top},
\end{equation}

where $N_{f}$ is the number of frames, $k_{r}$ is the number of frequency bins in $r$-th subband and $\bar{m}[k]$ is the mean subband power spectrum given by,

\begin{equation}\label{eqpia6}
\bar{m}[k]=\frac{1}{N_{f}}\sum_{i=1}^{N_{f}}P_{i}^{r}[k].
\end{equation}

\textbf{2. Computation of first PCA basis}: We apply \emph{singular value decomposition} (SVD)~\citep{golub2012matrix} to compute the PCA basis of subband covariance matrix. Using SVD, we can write, 

\begin{equation}\label{eqpia7}
\mathbf{S}^{r}=\mathbf{U}\mathbf{V}\mathbf{U}^{\top},
\end{equation}

where $\mathbf{U}$ is the $k_r \times k_r$ orthogonal matrix containing eigenvectors of $\mathbf{S}^{r}$ in each column, and the diagonal elements of the $k_r \times k_r$ matrix $\mathbf{V}$ contain the singular values. The first column of $\mathbf{U}$, i.e., the first principal component is used to create the $r$-th filter. We apply zero-padding to get the filter frequency response for the entire band. The computation of PCA-based filter response is illustrated in Fig.~\ref{fig:4}.

\begin{figure}[t!]
\centering
\includegraphics[width=0.9\textwidth]{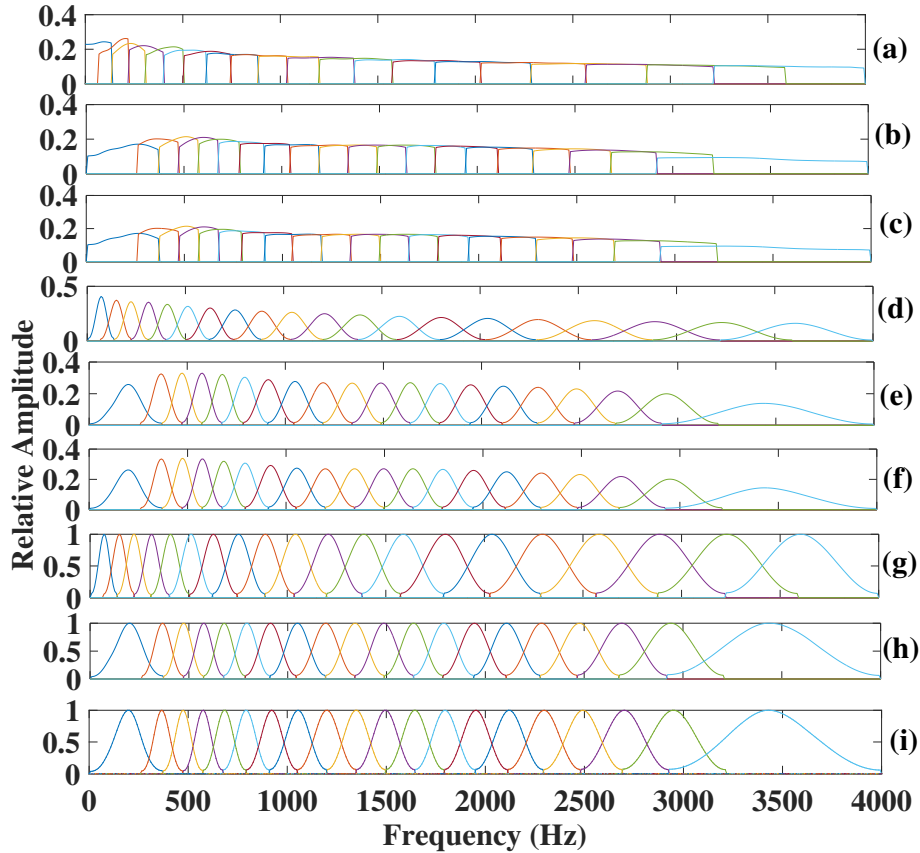}
\caption{Data-driven filterbank frequency responses using PCA. The filters are shown for different scales: (a) mel, (b) speech-based, and (c) speech-based with pitch. The next three (d, e and f) shows the filter shapes for three scales when Hamming window is applied on the log-power spectrum. The last three (g, h, and i) are for normalized frequency responses. In all cases, the filters are derived from the development set of NIST SRE $2001$ corpus.}\label{fig:5}
\centering
\end{figure}

The filter shape computed in the above process treats all the frequency components within a subband in an identical manner. However, considering the subbands have overlap with the adjacent bands, we apply tapering function to the power spectrum that assigns higher weights to the frequency components near center frequencies and lower weights to the components near edge frequencies. We use \emph{Hamming window} on the power spectrum data before performing PCA. We also~\textcolor{black}{normalize} the frequency response to make the highest magnitude unity similar to the mel filters. Fig.~\ref{fig:5} illustrates the filters for different frequency warping scale. 

In order to analyze the separability of different features, we compute F-ratio~\citep{nicholson1997}. For this analysis, we used $131$ speakers from POLYCOST corpus~\citep{hennebert2000polycost}. In Table~\ref{F-ratio}, we showed the F-ratio of log-energies of different feature extraction methods with $20$ filters. \textcolor{black}{This demonstrates that the proposed methods have more filters that have higher discriminative characteristics. We also showed the average F-ratio which indicates that the proposed features are better than the MFCC for most cases}.

\begin{table}[h]
\renewcommand{\arraystretch}{1.2}
\caption{\textcolor{black}{F-ratios of log-energies for MFCC features and for three kinds of SFCC features denoted by M1, M2 and M3}. M1 indicates the baseline SFCC feature where the scale is computed with all the speech frames. M2 indicates SFCC features when the scale is computed taking speech frames having pitch using pitch estimation algorithm. Finally, M3 indicates the SFCC features when the scale is same as M2 but triangular filters are replaced with window-based PCA filters. The last row shows the average ratio for all the cases.}
\centering
\begin{footnotesize}
\begin{tabular}{|l|c|c|c|c|}
\hline
\multirow{2}{*}{Filter No.}   & \multirow{2}{*}{MFCC} & \multicolumn{3}{|c|}{SFCC}\\ \cline{3-5}
           &                          &   M1              & M2                 & M3            \\ \hline
1          & $\textbf{0.5677}$        & $0.4107$          & $0.4103$           & $0.4149$       \\ \hline
2          & $\textbf{0.4241}$        & $0.3270$          & $0.3262$           & $0.3281$        \\ \hline
3          & $\textbf{0.3417}$        & $0.2864$          & $0.2862$           & $0.2864$         \\ \hline
4          & $0.2403$                 & $0.2855$          & $0.2856$           & $\textbf{0.2860}$ \\ \hline
5          & $0.2015$                 & $0.2965$          & $\textbf{0.2968}$  & $0.2961$         \\ \hline
6          & $0.2135$                 & $0.3018$          & $\textbf{0.3022}$  & $0.3012$         \\ \hline
7          & $0.2521$                 & $0.3209$          & $\textbf{0.3217}$  & $0.3209$          \\ \hline
8          & $0.2607$                 & $0.3405$          & $\textbf{0.3412}$  & $0.3410$           \\ \hline
9          & $0.2870$                 & $0.3564$          & $\textbf{0.3571}$  & $0.3565$          \\ \hline
10         & $0.3088$                 & $0.3773$          & $\textbf{0.3780}$  & $0.3774$           \\ \hline
11         & $0.3252$                 & $\textbf{0.3758}$ & $0.3753$           & $0.3749$             \\ \hline
12         & $0.3407$                 & $0.3775$          & $\textbf{0.3785}$  & $0.3781$              \\ \hline
13         & $0.3281$                 & $0.4093$          & $\textbf{0.4110}$  & $0.4110$               \\ \hline
14         & $0.3511$                 & $0.4396$          & $\textbf{0.4408}$  & $0.4404$                \\ \hline
15         & $0.3966$                 & $0.4596$          & $\textbf{0.4598}$  & $0.4593$                 \\ \hline
16         & $0.4280$                 & $\textbf{0.4469}$ & $0.4460$           & $0.4454$            \\ \hline
17         & $0.4160$                 & $0.4477$          & $\textbf{0.4487}$  & $0.4484$             \\ \hline
18         & $0.4557$                 & $0.4842$          & $0.4853$           & $\textbf{0.4861}$     \\ \hline
19         & $0.4990$                 & $0.5164$          & $\textbf{0.5176}$  & $0.5173$               \\ \hline
20         & $0.5696$                 & $0.5746$          & $0.5757$           & $\textbf{0.5845}$       \\ \hline
Avg.       & $0.3604$                 & $0.3917$          & $0.3922$           & $\textbf{0.3927}$        \\ \hline
\end{tabular}
\end{footnotesize}
\label{F-ratio}
\end{table}

\section{Experimental setup}\label{Section:Experimental setup}

\begin{table*}[h]
\renewcommand{\arraystretch}{1.2}
\caption{Summary of the corpora for speaker verification experiments.}
\centering
\begin{footnotesize}
\begin{tabular}{|c|c|c|c|c|c|c|}
\hline
\multirow{2}{*}{\textbf{Corpus}}&\textbf{No. of}&\textbf{Target}&\textbf{Test}&\textbf{Total}&\textbf{True}&\textbf{Impostor}\\
&\textbf{speakers}&\textbf{models}&\textbf{segments}&\textbf{trials}&\textbf{trials}&\textbf{trials}\\
\hline
NIST SRE $2001$&$174$&$174$&$2038$&$22418$&$2038$&$20380$ \\
\hline
NIST SRE $2002$&$330$&$330$&$3570$&$39270$&$2983$&$36287$\\
\hline
VoxCeleb1 &$40$&$4715$&$4713$&$37720$&$18860$&$18860$\\ 
\hline
\end{tabular}
\end{footnotesize}
\label{Database}
\end{table*}

\subsection{Corpora for experiments}
\par
We evaluated our proposed method in NIST (SRE $2001$ and SRE $2002$) and VoxCeleb (VoxCeleb1) speech corpora ~\citep{NIST1, NIST2, nagrani2017voxceleb1}. In addition, we evaluate the performance in noisy conditions. Initially, we conducted experiments on NIST SRE $2001$ corpus to~\textcolor{black}{optimize} different parameters. Then, we used those parameters to evaluate the ASV system in the NIST SRE $2002$ corpus for both clean and noisy conditions. 

We use VoxCeleb1 corpus consisting large number of speakers for real-world conditions~\citep{nagrani2017voxceleb1}. This corpus consists of voices of $1251$ celebrities collected from the YouTube videos. Out of them, $40$ speakers are used for evaluation purpose. The sampling rate of each utterance is of $16$ kHz, and average utterance length is $8$ seconds. The corpora used in our experiments are summarized in Table~\ref{Database}. 

We use the development data for scale computation. \textcolor{black}{The same data is used to train the model parameters and hyper-parameters, i.e., for computing the parameters for UBM, PLDA and T-matrix when required}. \textcolor{black}{The enrollment and the test sentences are corrupted with noises where SNRs range from $0$ to $40$ dB and type of noise is randomly chosen from five noises (white, pink, babble, volvo and factory). We took noise files from NOISEX-$92$ corpus.} 

\subsection{Feature extraction}
\par
We extracted the acoustic features from speech frames of $20$~ms with $10$~ms overlap. For experiments with GMM-UBM and i-vector system, we used $20$ filters. We extracted $19$ coefficients after discarding the first coefficient. Finally, a $57$-dimensional feature vector~\citep{sahidullah2016local} is formulated after appending delta and double-delta coefficients. The MFCCs are filtered with RASTA processing~\citep{hermansky1994rasta} to remove slowly varying channel effect. Finally, we perform cepstral mean and variance normalization (CMVN) after applying bi-Gaussian modelling based SAD~\cite{sahidullah2012design}. We use identical pre-processing and post-processing steps for all the features.

\subsection{Classifier details}

We use three different ASV systems: GMM-UBM, i-vector and DNN-based x-vector. \textcolor{black}{First, we use simple GMM-UBM classifier for conducting experiments with NIST SRE $2001$ and NIST SRE $2002$ corpora. Then, we evaluate our proposed feature on VoxCeleb1 corpus using i-vector and x-vector system. In order to make the work self-contained, we briefly describe all the classifiers as follows.}

\subsubsection{GMM-UBM system}
\textcolor{black}{In the GMM-UBM system, the feature distribution of the target speakers and the cohort models are represented with a mixture of Gaussian densities~\citep{reynolds2000speaker}. The cohort model, also known as universal background model (UBM) in this context, is trained with several hours of speech data}. The UBM is represented as $\mathbf{\lambda}_{\mathrm{ubm}}=\{ w_{i},\mathbf{\mu}_{i},\mathbf{\Sigma}_{i} \}_{i=1}^{C}$ where $C$ is the number of Gaussian mixture components, and $w_{i}$ is the weight, $\mathbf{\mu}_{i}$ is the mean, and $\mathbf{\Sigma}_{i}$ is the covariance matrix of the $i$-th component. The parameter $w_{i}$ satisfies the constrain $\textstyle \sum_{i=1}^{C}w_{i}=1$. The~\textcolor{black}{enrollment} speech model ($\lambda_{\mathrm{enroll}}$) are derived from the UBM using \emph{maximum-a-posteriori} (MAP) adaptation with the target speaker's feature.

During test, we calculate ASV score of the test utterance, $\mathbf{X}_{\mathrm{test}}= \{\mathbf{x}_{1}, \mathbf{x}_{2}, \ldots, \mathbf{x}_{T} \} $ as the log-likelihood ratio (LLR), given by the following equation:

\begin{equation}\label{eqpia10}
\Lambda_\mathrm{GMM-UBM}({\mathbf{X}_{\mathrm{test}},{\lambda_{\mathrm{enroll}}}})=\log P({\mathbf{X}_{\mathrm{test}}|\lambda_{\mathrm{enroll}}})-\log P({\mathbf{X}_{\mathrm{test}}|\lambda_{\mathrm{ubm}}}).
\end{equation}

Finally, if the ASV score is greater than or equal to a decision threshold, $\theta$, the test speech is considered as spoken by the correct speaker, otherwise an imposter.

In our experiments, we use the development section of NIST SRE $2001$ corpus, which consists of six hours of speech data, to train gender-independent UBM of $512$ mixture components. We use $10$ iterations of \emph{expectation-maximization} (EM) algorithm to estimate  the UBM parameters. The target speaker models are created by adapting only the mean vectors of UBM with relevance factor $14$.

\subsubsection{i-vector system}
In i-vector method, the GMM concatenated means of the adapted GMM, known as GMM-\emph{supervector}, is projected into a low dimensional space called as total variability (TV) space~\citep{dehak2010front} as,

\begin{equation}\label{eqpia11}
\mathbf{M}=\mathbf{m}+\mathbf{T}\mathbf{w},
\end{equation}

\textcolor{black}{where $\mathbf{T}$, $\mathbf{m}$ and $\mathbf{M}$ are the low-rank total variability matrix, the speaker and channel independent supervector (taken from UBM supervector) and the GMM supervector representation of the speech utterance, respectively}. Here the $\mathbf{w}$ is called as i-vectors. In order to compute the i-vector representation of a speech utterance, $\mathbf{X}_{\mathrm{utt}}$, we estimate the posterior mean of the i-vector given the centered first-order Baum-Welch statistics as,

\begin{equation}\label{eqpia12}
\mathbf{w}_{\mathrm{utt}}=(\mathbf{I}+ \mathbf{T}^{\top} \mathbf{\Sigma}^{-1}\mathbf{N}\mathbf{T})^{-1}\mathbf{T}^{\top}\mathbf{\Sigma}^{-1}\mathbf{F},
\end{equation}

\textcolor{black}{where $\mathbf{N}$ is matrix consisting of the zero-order Baum-Welch statistics as the diagonal elements; $\mathbf{F}$ is a vector whose elements are first-order Baum-Welch statistics;} and $\mathbf{\Sigma}$ is the residual variability, commonly created from the UBM covariances.

The extracted i-vectors contain channel information. \textcolor{black}{In order to compensate the effect of channel, \emph{probabilistic linear discriminant analysis} (PLDA) is used to compute the similarity between i-vectors of enrollment and test~\cite{rajan2014single}.} We use Gaussian PLDA (GPLDA) in our experiment which models the within-class covariance by a full-rank matrix.

The ASV score using PLDA is computed as the likelihood score given as,

\begin{equation}\label{eqpia13}
\Lambda_\mathrm{PLDA}(\mathbf{w}_{\mathrm{enroll}},{\mathbf{w}_{\mathrm{test}}})=\log\frac{p(\mathbf{w}_{\mathrm{enroll}},\mathbf{w}_{\mathrm{test}}|H_s)}{p(\mathbf{w}_{\mathrm{enroll}}|H_d)p(\mathbf{w}_{\mathrm{test}}|H_d)},
\end{equation}

where $\mathbf{w}_{\mathrm{enroll}}$ and $\mathbf{w}_{\mathrm{test}}$ are correspondingly the i-vectors of~\textcolor{black}{enrollment} and test sentences. Here $H_s$ and $H_d$ represent two hypotheses whether two i-vectors are from the same speaker ($H_s$) or not ($H_d$).

In our experiment with i-vector system, we have randomly selected $20,000$ and $50,000$ speech files from VoxCeleb1 dev set for training the UBM and T-matrix, respectively. The PLDA is trained with entire dev set consisting $148,642$ files from $1211$ speakers. We also apply linear discriminant analysis (LDA) to improve the speaker discriminativeness of i-vectors with the same data as used in PLDA training. We fix the number of mixture components to $512$ and i-vector dimension to $400$. The i-vectors are projected to $250$ dimensions using LDA. We perform \emph{whitening} and \emph{length normalization} on i-vectors before training GPLDA with $200$ dimensional speaker subspace.

\subsubsection{x-vector system}
~\textcolor{black}{The x-vector system uses deep neural network to learn the speech representation in a supervised manner unlike the unsupervised linear method used in i-vector approach~\citep{8461375}.} \textcolor{black}{The neural network consists of~\emph{time-delay neural network} (TDNN) along with statistical pooling followed by fully connected layers}. This architecture captures information from a large temporal context from the frame-level speech feature sequences~\citep{tdnn21701}. The TDNN is a fixed-size~\emph{convolutional neural network} (CNN) that share weights along the temporal dimension and it is regarded as 1D convolution (Conv1D) or temporal convolution~\citep{lecun1998gradient}. The x-vector system is trained for speaker classification task at segment level. Finally, the x-vectors are computed from the output of the first fully connected layer.

In our x-vector~\textcolor{black}{system} implementation, we use five TDNN layers and three fully connected layers as used in~\citep{8461375}. The details of the neural network configuration is shown in Table~\ref{xvector_description}.

\begin{table*}[!t]
\renewcommand{\arraystretch}{1.2}
\caption{Description of the layers in x-vector architecture.}
\centering
\vspace{0.1cm}
\begin{footnotesize}
\begin{tabular}{|c|c|}
\hline
\textbf{Layer}                   & \textbf{Details} \\
 \hline
TDNN-1                      & Conv1D (\#filter = 256, kernel size = 5, dilation rate =1)\\
 \hline
TDNN-2                      & Conv1D (\#filter = 256, kernel size = 3, dilation rate =2)\\
 \hline
TDNN-3                      & Conv1D (\#filter = 256, kernel size = 3, dilation rate =3)\\
 \hline
TDNN-4                      & Conv1D (\#filter = 256, kernel size = 1, dilation rate =1)\\
 \hline
TDNN-5                      & Conv1D (\#filter = 1024, kernel size = 1, dilation rate =1)\\
\hline
Statistics pooling & Computation of mean and standard deviation\\
\hline
FC1 & Fully connected layer (256 nodes) \\
\hline
FC2 & Fully connected layer (256 nodes) \\
\hline
Softmax & Softmax layer with 1211 outputs\\
\hline
\end{tabular}
\end{footnotesize}
\label{xvector_description}
\end{table*}

We implemented the x-vector system with Python library Keras~\citep{chollet2015keras} using TensorFlow~\citep{tensorflow2015-whitepaper} as backend. We use \emph{rectified linear unit} (ReLU)~\citep{nair2010rectified} and \emph{batch normalization}~\citep{ioffe2015batch} for all the five TDNN and two fully connected layers. We apply dropout with probability $0.05$ only on the two fully connected layers. The parameters of the neural network are initialized with Xavier normal method~\citep{glorot2010understanding}. The neural network is trained using Adam optimizer~\citep{kingma:adam} with learning rate $0.001$, $\beta_1=0.9$, $\beta_2=0.999$ and \textcolor{black}{without weight decay}. We train the neural network using speech segments of 1~seconds. We use $20$-dimensional MFCCs computed with $20$ filters. The MFCCs after dropping non-speech frames with SAD are processed with utterance-dependent cepstral mean~\textcolor{black}{normalization} (CMN). The x-vector systems are trained with batch size of $100$. We use the minimum validation loss as the stopping criteria. We consider entire VoxCeleb1 dev set consisting 1211 speakers (same data as i-vector extractor training). We used data augmentation as used in standard x-vector~\textcolor{black}{system}~\citep{8461375}. We extract $256$-dimensional embeddings from the fully connected layers (after batch~\textcolor{black}{normalization} but before applying ReLU).

\begin{table*}[h]
\renewcommand{\arraystretch}{1.2}
\caption{Parameters of the cost function for NIST SREs and VoxCeleb1 corpora.}
\centering
\vspace{0.1cm}
\begin{footnotesize}
\begin{tabular}{|c|c|c|c|}
\hline
Corpus                   & $C_{miss}$ &$C_{fa}$ & $P_{tar}$ \\
\hline
NIST SRE $2001$ and $2002$  & $10$ & $1$ & $0.01$               \\
\hline
VoxCeleb1                   & $1$ & $1$ & $0.01$                    \\
\hline
\end{tabular}
\end{footnotesize}
\label{costvalues}
\end{table*}

\subsection{Performance evaluation}
\textcolor{black}{We evaluate ASV system performance with commonly used evaluation metrics: \emph{equal error rate} (EER) and \emph{minimum detection cost function} (minDCF) computed from the detection error trade-off (DET) curve~\citep{kinnunen2010overview1, martin1997det}.} The EER is the point in DET curve where the false alarm rate (FAR) and false rejection rate (FRR) are equal. On the other hand, minDCF is computed by formulating a weighted cost function after assigning costs to the error rates followed by~\textcolor{black}{minimization} of the weighted cost function. The cost function is defined as,

\begin{equation}\label{eq_cost}
C_{det}= C_{miss} \times P_{miss} (\theta) \times P_{tar} +  C_{fa} \times P_{fa} (\theta) \times (1-P_{tar}),  
\end{equation}

where $C_{miss}$ and $C_{fa}$ are the cost of miss and false acceptance, respectively, $P_{miss}(\theta)$ and $P_{fa}(\theta)$ are the probabilities of miss and false acceptance at decision threshold $\theta$, and $P_{tar}$ is the prior target probability. The values of $C_{miss}$, $C_{fa}$ and $P_{tar}$ are chosen according to the evaluation plan of the respective corpus~\citep{nagrani2017voxceleb1, NIST1, NIST2} and their values are shown in Table~\ref{costvalues}.

\section{Results and discussion}\label{Section:Results and discussion}

\subsection{Experiments on NIST SREs with GMM-UBM system}
We evaluate the ASV performances on NIST SREs using GMM-UBM classifier. \textcolor{black}{First, we assess the performance with MFCC and baseline SFCC features for subsequent comparison with the proposed features on NIST SRE $2001$ corpus.} For SFCC methods, we compute the scale using the development section of NIST SRE $2001$ corpus. Table~\ref{Scores nist $2001$} shows the comparison between baseline MFCC, SFCC and proposed one (best one selected among pitch estimation methods mentioned) which indicates that the proposed one performs better than MFCC and SFCC in terms of both the evaluation metrics. 

\begin{table*}[h]
\renewcommand{\arraystretch}{1.2}
\caption{Comparison of ASV system performances in terms of EER (in \%) and minDCF$\times100$ for MFCC, SFCC, and the proposed features on NIST SRE $2001$ corpus using GMM-UBM backend.}
\centering
\vspace{0.1cm}
\begin{footnotesize}
\begin{tabular}{|c|c|c|c|}
\hline
\multicolumn{2}{|c|}{\textbf{Feature}} & \textbf{EER(in \%)} & \textbf{minDCF$\times100$} \\ \hline
\multicolumn{2}{|c|}{MFCC}        &  $7.70$   &   $3.39$     \\ \hline
\multicolumn{2}{|c|}{SFCC (Baseline)}        & $7.51$    &   $3.28$     \\ \hline
\multirow{5}{*}{SFCC (Scale with pitch)}& ~\citep{drugman2011joint}     &  $7.61$   &  $3.27$      \\ \cline{2-4} 
                       &~\citep{sun2002pitch}     &  $7.45$   &   $3.40$     \\ \cline{2-4} 
                       &~\citep{de2002yin}     &  $7.31$   &  $\textbf{3.23}$      \\ \cline{2-4} 
                       &~\citep{wise1976maximum}     & $7.22$    &   $3.26$     \\ \cline{2-4} 
                       &~\citep{gonzalez2014pefac}     & $\textbf{7.21}$    &  $3.24$      \\ \hline
\end{tabular}
\end{footnotesize}
\label{Scores nist $2001$}
\end{table*}

We also perform the experiment with data-driven filter shapes created with PCA-based method. In Table~\ref{data_driven_filter_PCA_only}, we have shown the ASV performance for different scales where the filter is computed with PCA on the development data. We observe that the ASV performance is relatively poor compared to the results with fixed triangular based filters. Interestingly, the proposed scale based features are better than mel scale based features. The pitch based ASV system yields lowest EER amongst all the three systems. 

\begin{table*}[t!]
\renewcommand{\arraystretch}{1.2}
\caption{Comparison of ASV performances with PCA-based data-driven filters using different scales. Results are shown in terms of EER (in \%) and minDCF$\times100$ on NIST SRE $2001$ corpus using GMM-UBM back-end.}
 \centering
  \vspace{0.1cm}
  \begin{footnotesize}
\begin{tabular}{|c|c|c|}
\hline
\textbf{Scale}  & \textbf{EER(in \%)} & \textbf{minDCF$\times100$}   \\
\hline
Mel                            &$8.69$&$3.86$        \\
\hline
Speech-based                   & $8.39$&$\textbf{3.61}$ \\
\hline
Speech-based with pitch        &$\textbf{8.38}$&$3.66$     \\
\hline
\end{tabular}
\label{data_driven_filter_PCA_only}
\end{footnotesize}
\end{table*}

We further apply tapering window (here Hamming) on the subband spectrum before performing PCA. The results are reported in Table~\ref{data_driven_filter_PCA_symmetric_window}. We observe noticeable improvement compared to the results of untaperd case in Table~\ref{data_driven_filter_PCA_only}. Interestingly, the performance obtained with the data-driven filter shapes are sometimes better than the performance with triangular filters. For instance, in case of MFCCs, the minDCFs ($\times$ 100) of the triangular and data-driven filters are $3.39$ (Table~\ref{Scores nist $2001$}) and $3.35$ (Table~\ref{data_driven_filter_PCA_symmetric_window}). Similarly, the EER for pitch-based system reduces to $7.11\%$ from $7.21\%$ when windowed and PCA-based data-driven filter is used instead of triangular filters. However, we do not observe improvement in EER with after~\textcolor{black}{normalizing} the filter response magnitudes though we observe a reduction in cost function values.

\begin{table*}[t!]
\renewcommand{\arraystretch}{1.2}
\caption{Same as Table~\ref{data_driven_filter_PCA_only} but with tapering Window applied on the subbands before applying PCA. We also report the performance with magnitude~\textcolor{black}{normalized} filters in the last row.}
 \centering
  \vspace{0.1cm}
  \begin{footnotesize}
\begin{tabular}{|c|c|c|}
\hline
\textbf{Scale}  & \textbf{EER(in \%)} & \textbf{minDCF$\times100$} \\
\hline
Mel                                        &$7.70$&$3.35$    \\
\hline
Speech-based                               &$7.25$&$3.27$       \\
\hline
Speech-based with pitch                    &$\textbf{7.11}$&$3.23$\\
\hline
Speech-based with pitch (Normalized)       &$7.41$&$\textbf{3.22}$  \\
\hline
\end{tabular}
\label{data_driven_filter_PCA_symmetric_window}
\end{footnotesize}
\end{table*}

\begin{table*}[t!]
\renewcommand{\arraystretch}{1.2}
\caption{Comparison of ASV performances with fixed (i.e, mel scale with triangular filter) and various data-driven features on NIST SRE $2002$ corpus. Performances are shown in terms of EER (in \%) and minDCF $\times100$ using GMM-UBM backend. \textcolor{black}{Here, the scale is computed using development set of NIST SRE 2001 corpus}.}
\centering
 \vspace{0.1cm}
 \begin{footnotesize}
\begin{tabular}{|c|c|c|c|}
\hline
\textbf{Filter Shape} & \textbf{Scale} & \textbf{EER (in \%)} & \textbf{minDCF $\times100$}  \\ \hline
\multirow{3}{*}{Triangular}     & Mel &$8.76$&$4.07$   
                                \\ \cline{2-4} 
                                                                            
                                & Speech-based  &$9.15$&$4.45$      
                                \\ \cline{2-4} 
                                                                                
                                & Speech-based with pitch &$9.12$  &$4.28$  
                                \\ \hline
\multirow{3}{*}{PCA}                 & Mel  &$9.65$&$4.33$                                                           \\ \cline{2-4} 
                                    & Speech-based  &$9.92$&$4.55$ 
                                    \\ \cline{2-4} 
                                    & Speech-based with pitch  &$9.96$&$4.57$   \\ \hline
\multirow{3}{*}{Window+PCA}         & Mel  &$\textbf{8.42}$&$4.04$   
                                   \\ \cline{2-4} 
                                                                           
                                   & Speech-based  &$9.15$&$4.34$       
                                   \\ \cline{2-4} 
                                                                                 & Speech-based with pitch  &$8.75$&$4.25$   \\ \hline
\multirow{3}{*}{Window+PCA+Norm.}   & Mel  &$8.48$&$\textbf{4.03}$  
                                    \\ \cline{2-4} 
                                                                           & Speech-based&$9.29$&$4.29$      
                                    \\ \cline{2-4} 
                                                                           & Speech-based with pitch  &$8.91$&$4.33$ \\ \hline
\end{tabular}
 \label{results NIST $2002$ database}
 \end{footnotesize}
\end{table*}

\begin{figure}[t!]
\begin{center}
\includegraphics[width=1\textwidth,trim={9cm 0 8cm 0cm},clip]{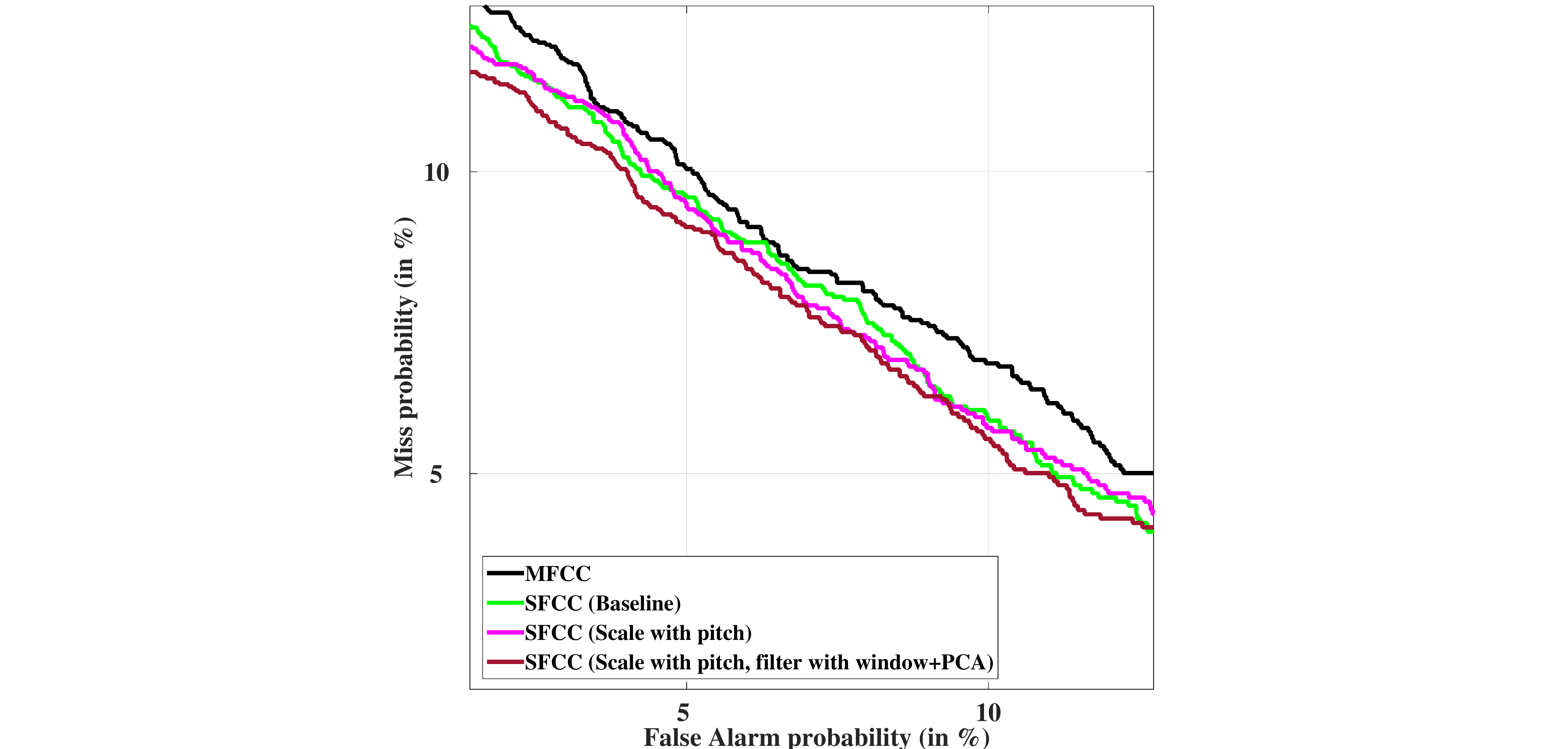}
\end{center}
\vspace{-.5cm}
\caption{The DET curves ASV system performance using different feature extraction methods on NIST SRE $2001$ corpus with GMM-UBM as backend.}\label{fig:6}
\centering
\vspace{-0.25cm}
\end{figure}

\begin{table*}[t!]
\renewcommand{\arraystretch}{1.2}
\caption{Comparison of ASV system performances in noisy conditions. Results are shown in terms of EER (in \%) and minDCF$\times100$ on additive noise-corrupted NIST SRE $2002$ corpus with GMM-UBM as backend.}
 \centering
  \vspace{0.1cm}
  \begin{footnotesize}
\begin{tabular}{|c|c|c|}
\hline
\textbf{Methods}  & \textbf{EER(in \%)} & \textbf{minDCF$\times100$}           \\
\hline
MFCC (Baseline)                                             &$18.27$&$8.04$       \\
\hline
Speech-based with triangular filter                         &$16.63$&$7.64$          \\
\hline
Speech-based (pitch) with window \& PCA-based filter  &$\textbf{16.02}$&$\textbf{7.56}$    \\
\hline
\end{tabular}
\label{noisy_2002}
\end{footnotesize}
\end{table*}

\begin{figure}[t!]
\begin{center}
\includegraphics[width=1\textwidth,trim={10cm 0 8cm 0cm},clip]{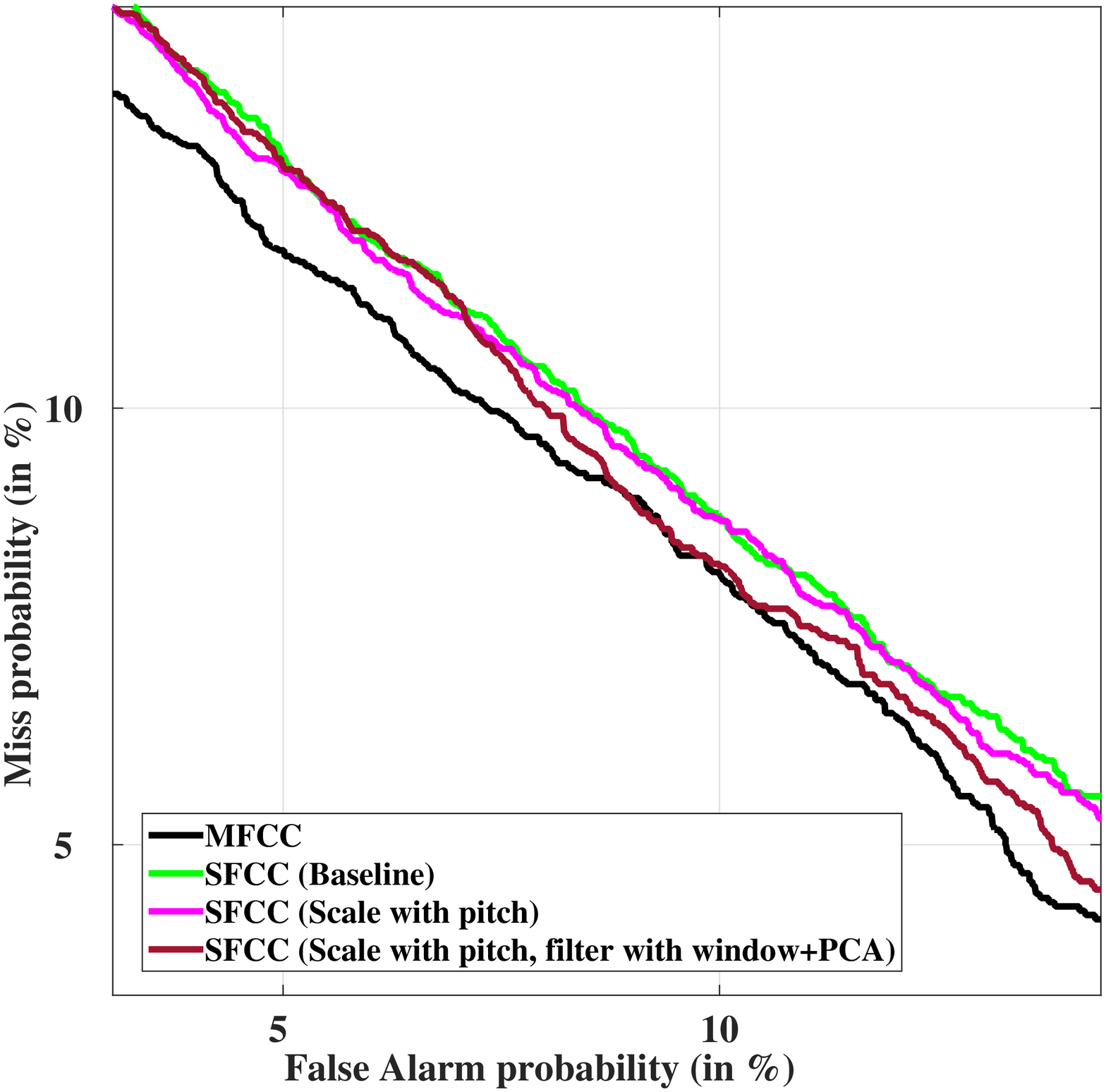}
\end{center}
\vspace{-.5cm}
\caption{Same as Fig.\ref{fig:6} but for NIST SRE $2002$ corpus.}\label{fig:7}
\centering
\vspace{-.5cm}
\end{figure}

We also conduct experiment with NIST SRE $2002$ corpus to evaluate the~\textcolor{black}{generalization} ability of the proposed data-driven approach. In this case, the same development data from the subset of NIST SRE $2001$ corpus is used for computing the parameters of data-driven feature extractor. The results are summarized in Table~\ref{results NIST $2002$ database}. \textcolor{black}{We observe that with triangular filter, mel-scaled filterbank always obtain lower EER and minDCF than the data-driven scale based methods. The reason for this performance is due to domain-mismatch as the scale is computed on the speech files from a different corpus, i.e., NIST SRE 2001.} However, we notice that the warping scale based on the selected frames from pitch show improvement over the condition where the scale is computed on all the speech frames (\emph{i.e.}, baseline SFCCs). The DET curves of ASV results of selected features are illustrated in Fig.~\ref{fig:6} and~\ref{fig:7}.

\begin{figure}[t!]
\centering
\includegraphics[width=1\textwidth,trim={10cm 0 8cm 0cm},clip]{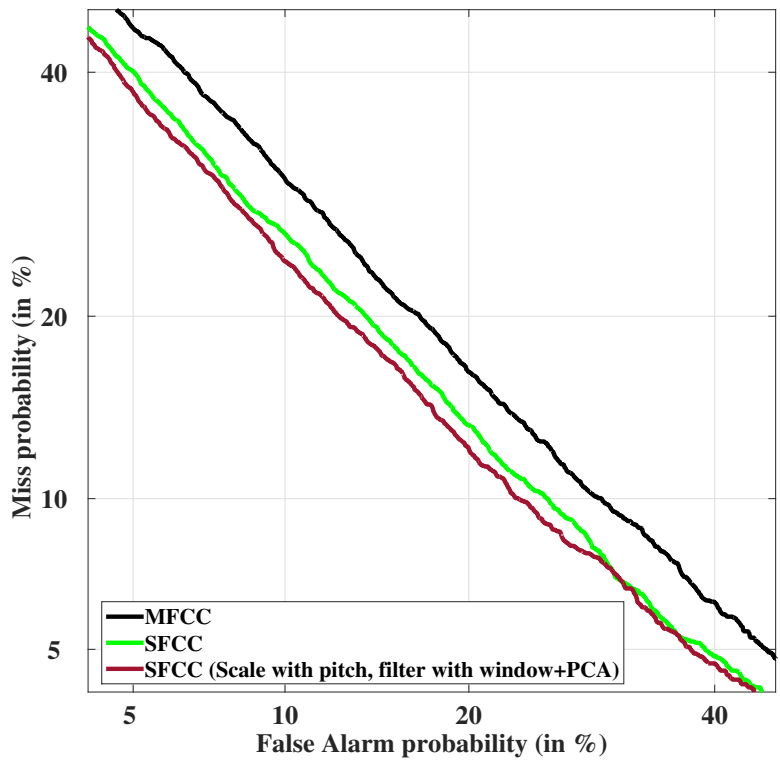}
\caption{The DET curves of ASV systems based on different feature extraction methods on noise-corrupted version of NIST SRE $2002$ corpus using GMM-UBM as backend. }\label{fig:8}
\vspace{-0.5cm}
\end{figure}

\begin{figure}[t!]
\centering
\includegraphics[width=1\textwidth,trim={10cm 0 8cm 0cm},clip]{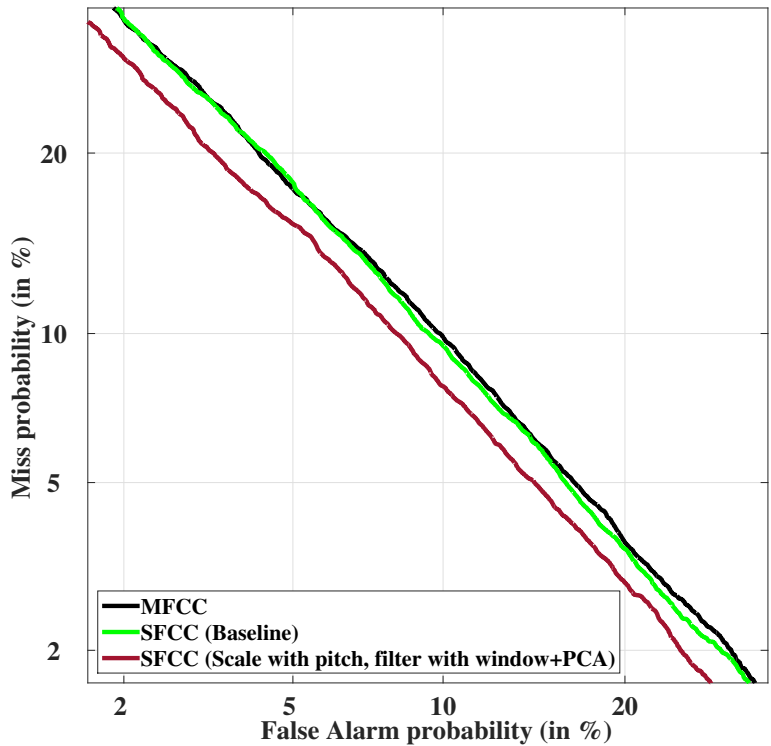}
\caption{The DET curves ASV systems based on different data-driven feature extraction methods on VoxCeleb1 corpus with i-vector and PLDA-based scoring as backend.}
\label{fig:9}
\vspace{-0.25cm}
\end{figure}

\begin{table*}[t!]
\renewcommand{\arraystretch}{1.2}
\caption{Comparison of ASV performances on VoxCeleb1 corpus with i-vector system. Results are shown in terms of EER (in \%) and minDCF$\times100$ for features based on different scales. \textcolor{black}{The scale is computed on the development set of the VoxCeleb1 corpus.}}
\centering
\vspace{0.1cm}
\begin{footnotesize}
\begin{tabular}{|c|c|c|}
\hline
\textbf{Scale}  & \textbf{EER(in \%)} & \textbf{minDCF$\times100$}   \\
\hline
Mel                     &$9.95$&$0.747$             \\
\hline
Speech-based            &$9.71$&$0.786$                \\
\hline
Speech-based with pitch &$9.52$&$\textbf{0.721}$ \\
\hline
Speech-based (pitch) with window \& PCA-based filter &$\textbf{8.98}$&$0.744$ \\
\hline
\end{tabular}
\label{voxceleb1_results}
\end{footnotesize}
\end{table*}

\begin{figure}[t!]
\centering
\includegraphics[width=.9\textwidth,trim={.75cm 0 .5cm 0cm},clip]{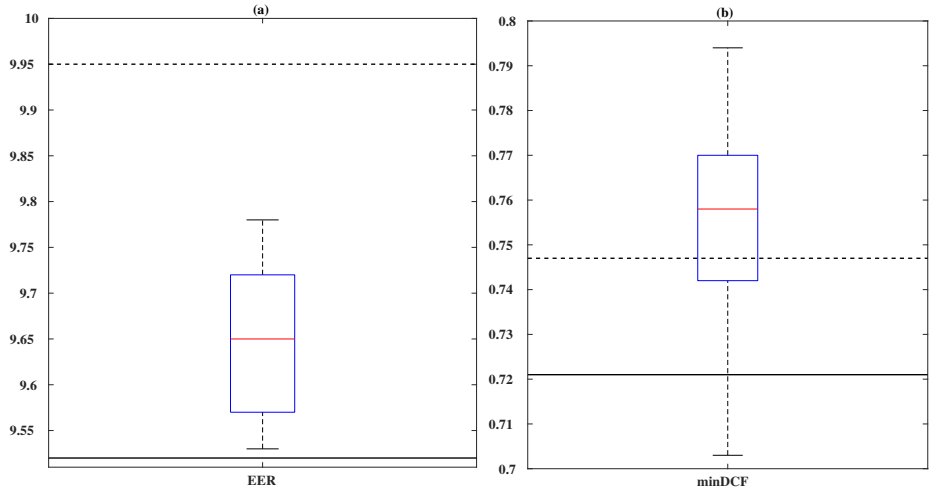}
\caption{Error bar plot showing ASV performance on VoxCeleb1 where 0.1\% of the total speech data are randomly selected for computing filterbank parameters. The results are shown on VoxCeleb1 corpus with i-vector back-end. The dotted horizontal line indicates the performance with baseline MFCCs and continuous horizontal line denotes the performance with proposed method where 100\% speech data are used for computing the filterbank parameters.}
\label{fig:10}
\end{figure}
\begin{table*}[t!]
\renewcommand{\arraystretch}{1.2}
\caption{Comparison of ASV performances when in-domain and out of domain (Librispeech and TIMIT) are used for computing scale of data-driven filter. Results are shown in terms of EER (in \%) and minDCF$\times100$ on VoxCeleb1 test set using i-vector and PLDA backend.}
\centering
\vspace{0.1cm}
\begin{footnotesize}
\begin{tabular}{|c|c|c|}
\hline
\textbf{Corpus for scale computation} & \textbf{EER(in \%)} & \textbf{minDCF$\times100$}   \\
\hline
VoxCeleb1 (in-domain)      & $\textbf{9.52}$      & $\textbf{0.721}$        \\
\hline
Librispeech    &    $10.40$   &    $0.812$    \\
\hline
TIMIT        & $9.99$      &     $0.730$    \\
\hline
\end{tabular}
\end{footnotesize}
\label{domain_scale_voxceleb1_results}
\end{table*}

Even though we do not observe improvement with the data-driven scales, the performances of mel scale based are improved with window and PCA based data-driven filters. We can conclude that scale selection is more sensitive to the corpus selection whereas filter-responses computed from one dataset generalize well to other datasets.

Finally, the results on noisy conditions are shown in Table~\ref{noisy_2002} and the corresponding DET in Fig.~\ref{fig:8}. Here, we have found that the propose data-driven features are more robust compared to the baseline MFCCs. The best performance in terms of EER is obtained with data-driven feature where scale is computed from the selected frames with pitch values and the filter shape is computed with windowed spectrum and PCA.

\subsection{Experiments on VoxCeleb1}

\subsubsection{Performance evaluation with i-vector system}

In our experiments with i-vector system on VoxCeleb1, first we compute the scale on the entire development set consisting of $1211$ speakers and report the results for different scales in Table~\ref{voxceleb1_results}. \textcolor{black}{We observe that the performance with feature using frame selection based scale yields better performance in terms of both EER and minDCF.} We obtain more than $4.30\%$ and $3.48\%$ relative reduction for EER and minDCF, respectively. Fig.~\ref{fig:9} shows the DET curve of the ASV system using VoxCeleb1 corpus.~\textcolor{black}{From this curve, we find that the proposed features perform better than the other features in ASV task}.

\begin{table*}[t!]
\renewcommand{\arraystretch}{1.2}
\caption{Comparison of ASV performances with x-vector representation. Results are shown in terms of EER (in \%) and minDCF$\times100$ on VoxCeleb1 test set.}
\centering
\vspace{0.1cm}
\begin{footnotesize}
\begin{tabular}{|c|c|c|c|c|}
\hline
\multirow{2}{*}{\textbf{Methods}} & \multicolumn{2}{|c|}{\textbf{Embeddings from FC1}} & \multicolumn{2}{|c|}{\textbf{Embeddings from FC2}} \\ \cline{2-5} 
                & \textbf{EER (in \%)} & \textbf{minDCF$\times100$} & \textbf{EER (in \%)} & \textbf{minDCF$\times100$} \\ 
\hline
MFCC                 & $5.13$ & $0.468$  & $5.19$ & $0.480$             \\ 
\hline    
Proposed SFCC        & $5.03$ & $0.468$ & $4.96$ & $0.502$             \\ 
\hline    
Score Fusion        & $\textbf{4.45}$  & $\textbf{0.421}$  & $\textbf{4.56}$ 
                    &  $\textbf{0.446}$           \\ 
\hline
\end{tabular}
\end{footnotesize}
\label{x_vector_voxceleb1_results}
\end{table*}

\begin{figure}[t!]
\centering
\includegraphics[width=1\textwidth,trim={.5cm 0 .5cm 0cm},clip]{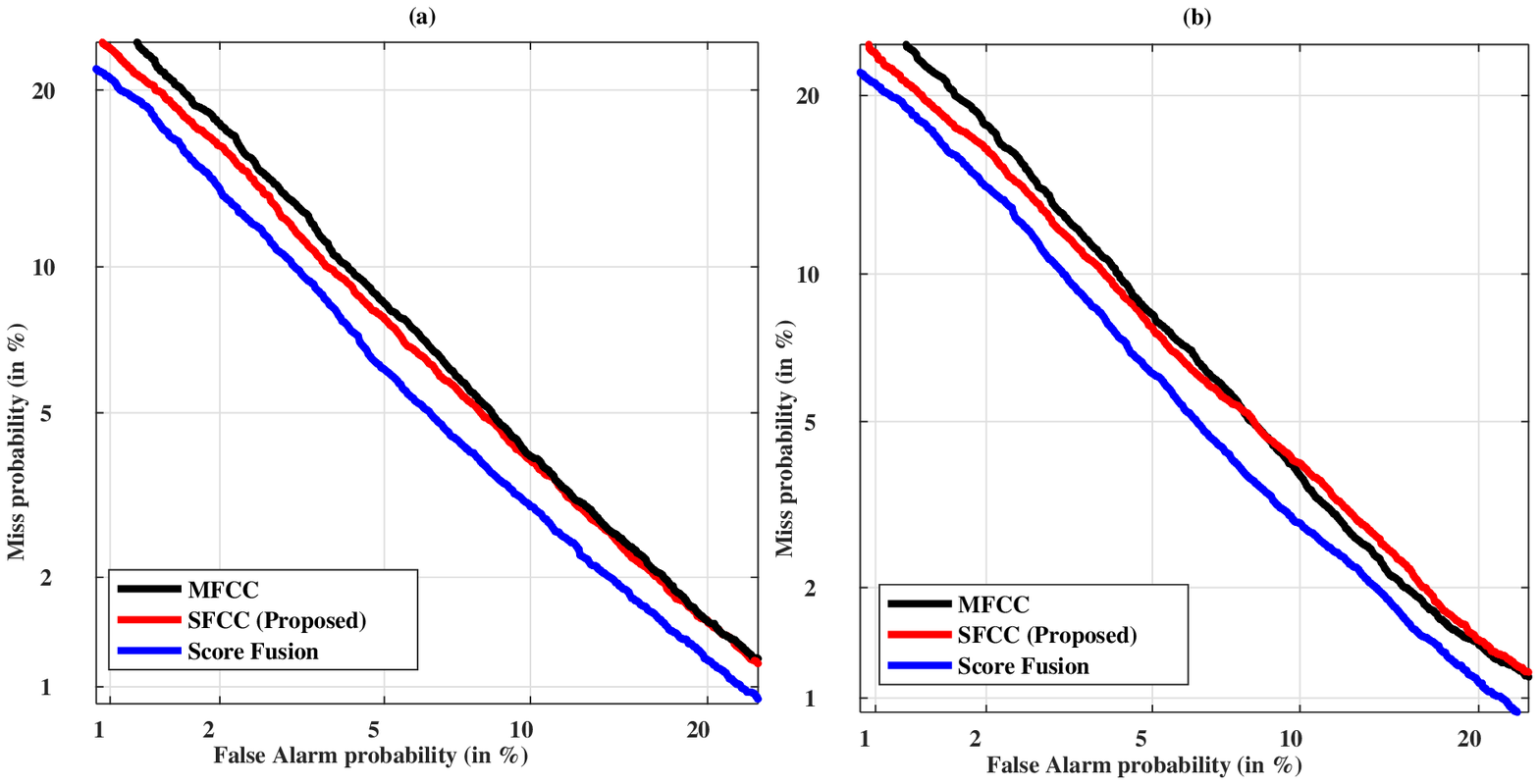}
\vspace{-3cm}
\caption{The DET curves of results of ASV system among MFCC, Proposed feature extraction methods and score level fusion on VoxCeleb1 corpus using x-vector system where (a) using FC1 and (b) using FC2. }
\label{fig:11}
\end{figure}

In Table~\ref{voxceleb1_results}, the scales are computed with entire development data which is computationally expensive, especially for PCA-based filter shape computation. In the next experiment, we examined the effect of amount of data for scale computation on the performance of ASV system where we chose a small subset of speech utterances from the entire set of $148642$ files. We conducted the ASV experiments 10 times where every time 0.1\% of the speech data are randomly selected. The ASV performances for this randomly chosen small subsets are shown in Fig.~\ref{fig:10}. The figure also shows the performance with baseline MFCCs and proposed method with full speech data. We observe that filterbank parameters computed with 0.1\% of the data shows lower EER than baseline MFCCs. However, we do not observe improvements in minDCF. Interestingly, the ASV performance with 100\% of the data for scale computation only gives about 1\% relative improvement (in terms of EER) over 0.1\% data. To compare the performance with out of domain data, we also conduct experiment where the data for scale formulation is taken from corpus other than VoxCeleb1 in-domain data. We took speech data from Librispeech~\citep{panayotov2015librispeech} and TIMIT~\citep{zue1990speech} for this purpose. We use the same VoxCeleb1 data for computing parameters of UBM, T-matrix, LDA and PLDA. The results reported in Table~\ref{domain_scale_voxceleb1_results} indicates that out of domain data considerably degrades the ASV performances. We conclude that the proposed method should be applicable where limited in-domain data is available.

\subsubsection{Performance evaluation with x-vector system}
For experiments with x-vector system, we chose baseline MFCCs and the proposed data-driven feature extraction in which the warping scale and the filter parameters are computed with development data from the VoxCeleb1 corpus. The results of x-vector~\textcolor{black}{system} with PLDA scoring are~\textcolor{black}{summarized} in Table~\ref{x_vector_voxceleb1_results}. In the state-of-the-art x-vector system also, the proposed features are better than conventionally used MFCCs in terms of EER. We showed the results when the embeddings are computed from the output of FC1 and FC2. The improvement is observed for both cases. We did not find improvement in terms of minDCF; however, the proposed features are better than MFCCs in most of the operating points as shown in the DET curve in Fig.\ref{fig:11}.

Finally, we perform experiments with fused system where scores of MFCC and proposed SFCC are combined with equal weights~\citep{6494266}. The performance is substantially improved with fusion. In EER, we obtained relative improvement of $13.26\%$ and $12.14\%$ over baseline MFCC, respectively for x-vector embeddings computed from FC1 and FC2. This confirms the complementarity of proposed data-driven filterbank with mel filterbank.

\section{Conclusion}\label{Section:Conclusion}
\textcolor{black}{The filterbanks in most of the acoustic feature extraction modules are either handcrafted with some auditory knowledge or learned over a large dataset with some objectives. In this work, we proposed to compute the MFCC filterbank in a data-driven way. We improved the data-driven frequency warping scale by considering voiced frames having pitch information. We demonstrated the superiority of the newly designed warping scale for ASV tasks. We also computed frequency responses of the filters in a data-driven manner from the subband power spectrum using PCA. We showed that both these schemes reduce the speaker recognition error rates. We observed improvements in both matched and mismatch conditions. The proposed feature extraction method is compatible with the state-of-the-art x-vector systems and shows improvement over MFCC-based ASV systems. The proposed method is computationally less expensive than DNN-based data-driven methods. Also, it computes the filterbank parameters (\emph{i.e.}, filter edge frequencies \& frequency response) with a small amount of speech data without additional metadata as opposed to the supervised methods which require a large amount of labeled data. We further improved the ASV performance by simple score fusion with an MFCC-based system.}


\textcolor{black}{Even though the acoustic features computed with the proposed data-driven filters show improvement over MFCCs, the performance of the proposed features substantially degrades if in-domain audio-data is not available. However, domain-mismatch remains an open challenge for other data-driven feature extractors, too. In future, we plan to explore the data-augmentation methods for addressing the domain mismatch issue. We can compute the filterbank from the augmented speech data and observe its robustness. The objective of this work was not to optimize the number of filters and the amount of overlaps with the adjacent filters. The present work can also be extended in that direction. In this work, we develop the filterbank in a task-independent manner but its application is limited to ASV in the current study. We also plan to adopt the proposed data-driven filterbank for other potential speech processing tasks, such as language and emotion recognition.}

\section*{Acknowledgments}
The authors would like to express their sincere thanks to the anonymous reviewers and the editors for their valuable comments and suggestions which greatly improved the work in quality and content. The work of Md Sahidullah is supported by Region Grand Est, France. Experiments presented in this paper were partially carried out using the Grid'5000 testbed, supported by a scientific interest group hosted by Inria and including CNRS, RENATER and several Universities as well as other organizations (see \url{https://www.grid5000.fr}).

\bibliography{latexbibp.bib}
\section*{}

\textbf{Susanta Sarangi} is currently pursuing Ph.D. in area of speech processing from the Department of Electronics \& Electrical Communication Engineering, Indian Institute of Technology Kharagpur. He obtained Master of Technology degree in Electronics and Communication Engineering  from Biju Patnaik University of Technology, Odisha in 2008 and Bachelors of Engineering in Electronics and Telecommunication Engineering from Utkal University, Odisha in 2002. Before joining Ph.D., he was an Assistant Professor in Institute of Technical Education and Research (ITER), Siksha 'O' Anusandhan Deemed to be University, Odisha. His research interests include speech \& audio signal processing, signal processing, and machine learning.

\vspace{1cm}

\textbf{Md Sahidullah} received his Ph.D. degree in the area of speech processing from the Department of Electronics \& Electrical Communication Engineering, Indian Institute of Technology Kharagpur in 2015. Prior to that he obtained the Bachelors of Engineering degree in Electronics and Communication Engineering from Vidyasagar University in 2004 and the Masters of Engineering degree in Computer Science and Engineering from West Bengal University of Technology in 2006. In 2014-2017, he was a postdoctoral researcher with the School of Computing, University of Eastern Finland. In January 2018, he joined MULTISPEECH team, Inria, France as a post-doctoral researcher where he currently holds a starting research position. His research interest includes robust speaker recognition and spoofing countermeasures. He is also part of the organizing team of two Automatic Speaker Verification Spoofing and Countermeasures Challenges: ASVspoof 2017 and ASVspoof 2019. Presently, he is also serving as Associate Editor for the IET Signal Processing and Circuits, Systems, and Signal Processing.

\vspace{1cm}

\textbf{Goutam Saha} received his B.Tech. and Ph.D. degrees from the Department of Electronics \& Electrical Communication Engineering, Indian Institute of Technology (IIT) Kharagpur, India in 1990 and 2000, respectively. In between, he served industry for about four years and obtained a five year fellowship from Council of Scientific \& Industrial Research, India. In 2002, he joined IIT Kharagpur as a faculty member where he is currently serving as a Professor. His research interests include analysis of audio and bio signals.

\end{document}